%% file: main.tex
\documentclass{article}

\usepackage{spconf}

\usepackage[utf8]{inputenc} 
\usepackage[T1]{fontenc}    
\usepackage{hyperref}       
\usepackage{url}            
\usepackage{booktabs}       
\usepackage{amsfonts}       
\usepackage{nicefrac}       
\usepackage{microtype}      
\usepackage{xcolor}         

\usepackage{amsmath}
\usepackage{amssymb}
\usepackage{bbm}
\usepackage{siunitx}
\usepackage{graphicx}
\usepackage[inkscapelatex=false]{svg}
\usepackage{csvsimple-legacy}
\usepackage{pgfplotstable}
\usepackage{makecell}
\usepackage{multirow}
\usepackage{subfigure}
\usepackage{enumitem}
\usepackage{wrapfig}
\usepackage{caption}

\newcommand{\mdprnn}{Multi-channel Dual-path RNN}

\newcommand{\ours}{\texttt{FEABASE}}

\newcommand{\setD}{\mathcal{D}}
\newcommand{\setDall}{\setD{}_\textnormal{all}}
\newcommand{\setDeasy}{\setD{}_\textnormal{easy}}
\newcommand{\setDhard}{\setD{}_\textnormal{hard}}

\newcommand{\setS}{\mathcal{S}}
\newcommand{\setSeasy}{\setS{}_\textnormal{easy}}
\newcommand{\setShard}{\setS{}_\textnormal{hard}}
\newcommand{\setSboth}{\setS{}_\textnormal{easy+hard}}

\newcommand{\setTeasy}{\Tilde{\setD{}}_\textnormal{easy}}
\newcommand{\setThard}{\Tilde{\setD{}}_\textnormal{hard}}
\newcommand{\setTall}{\Tilde{\setD{}}_\textnormal{all}}

\renewcommand{\paragraph}[1]{{\noindent \textbf{ #1. }}}

\definecolor{myblue}{RGB}{0, 0, 255} 
\definecolor{mydarkblue}{RGB}{0, 0, 139} 
\hypersetup{
    colorlinks=true,
    linkcolor=myblue,
    filecolor=myblue,
    citecolor=myblue,
    urlcolor=myblue,
    linkbordercolor=mydarkblue,
    filebordercolor=mydarkblue,
    citebordercolor=mydarkblue,
    urlbordercolor=mydarkblue,
}

\title{
Addressing feature imbalance in sound source separation
}

\name{Jaechang Kim$^1$,  Jeongyeon Hwang$^1$, Soheun Yi$^2$, Jaewoong Cho$^2$, Jungseul Ok$^1$}
\address{$^1$POSTECH, $^2$KRAFTON 
}

\begin{document}

\maketitle

\begin{abstract}

Neural networks often suffer from a feature preference problem, where they tend to overly rely on specific features to solve a task while disregarding other features, even if those neglected features are essential for the task. 
Feature preference problems have primarily been investigated in classification task. However, we observe that
feature preference occurs in high-dimensional regression task, specifically, source separation.
To mitigate feature preference in source separation, we propose \emph{FEAture BAlancing by Suppressing Easy feature \texttt{(FEABASE)}}. 
This approach enables efficient data utilization by learning hidden information about the neglected feature.
We evaluate our method in a multi-channel source separation task, where feature preference between spatial feature and timbre feature appears.

\end{abstract}


\section{Introduction}
\label{sec:intro}

\input{intro}

\input{problem}

\input{method}

\input{experiments}

\vspace{-1mm}
\section{Conclusion}
\vspace{-1mm}
\label{sec:conclusion}

We address the imbalanced feature preference in 
sound source separation, which is a fundamental task in signal processing.
This is a first discovery of the imbalanced problem for high-dimensional regression task.
We proposed \ours{} to resolve the feature imbalance.
Our experiment shows that the proposed method indeed enables learning
both easy and hard features, which was infeasible in the standard learning.
We believe that the proposed framework can be extended to other domains and 

\bibliography{ref}

\clearpage

\end{document}

%% file: intro.tex
Despite remarkable advances in deep learning,
standard deep learning often has an imbalanced feature preference, which induces the trained neural network to overly rely on features easy to learn
and to ignore hard yet crucial features.
This is also known as simplicity bias~\cite{simplicity}
or shortcut learning~\cite{minderer2020shortcut}, and causes
poor generalization mainly due to the biased inference towards only the easy features, e.g., given a waterbird image taken on river, identifying the species based on the feature of background rather than that of bird.

The imbalance problem has been addressed
in various contexts: debiasing~\cite{nam2020LfF, jtt}
and long-tailed learning~\cite{kim2020m2m}.
The previous works have focused on simple classification tasks.
However, we discover and address the imbalanced feature preference in sound source separation, generating source signals from a mixture signal. This is particularly interesting
since it is one of fundamental tasks in signal processing
and also it is a high-dimensional regression task, where a neural network
is trained to generate output signals containing features as detailed as the input is, whereas a classifier maps high-dimensional input to low-dimensional output.

To be specific, we consider deep learning for a multi-channel source separation (MSS) \cite{coneofsilence, fasnettac}, where the mixture is recorded using multiple microphones.
In single-channel source separation, the timbre of the source signal is a key feature.
In addition to timbre feature, multi-channel sound have spatial feature.
The spatial characteristic of sources, such as
the direction of arrival, is also a prominent feature~\cite{doa}.
However, 
in Section~\ref{sec:problem},
we found that standard deep learning for MSS is prone to preferring only the spatial feature. In other words, the easy and hard features
of MSS are the spatial and timbre features, respectively.
Such an imbalanced feature preference indeed leads to fatal failures, particularly when only the timbre feature is available, e.g., (b) in Figure~\ref{fig:multi_sep_setting}.

\input{figures/figure_multi_sep_setting}

To tackle the problem of imbalanced feature preference in MSS,
we aim at suppressing the easy feature in training a neural network
so that it is encouraged to learn the hard feature, while learning the easy feature, of course.
To this end, in Section~\ref{sec:method}, we devise a 
feature balancing method by suppressing easy feature, called {\tt FEABASE},
which augments
training samples by suppressing the easy feature, 
and then train a neural network with the augmented dataset
of a balanced feature preference.
In Section~\ref{sec:experiment}, we demonstrate a balanced feature learning 
for MSS using the proposed method.

    To summarize our contribution,
\begin{itemize}
\vspace{-1mm}
\item We discover the feature preference problem regarding a high-dimensional regression task, and thoroughly analyze the feature preference problem in source separation (Section~\ref{sec:problem}).
\vspace{-1mm}

\item We propose \texttt{FEABASE} which balances the feature preference by suppressing the easy feature in training samples, while maintaining the informativeness of the given dataset as much as possible (Section~\ref{sec:method}).
\vspace{-1mm}

\item We demonstrate the efficacy of the proposed method given a canonical dataset of MSS. It indeed enables learning both of the timbre and spatial features, which is not available by the standard deep learning.

\end{itemize}

%% file: figures/figure_multi_sep_setting.tex
\begin{figure}[t]
    \includegraphics[width=1\columnwidth]{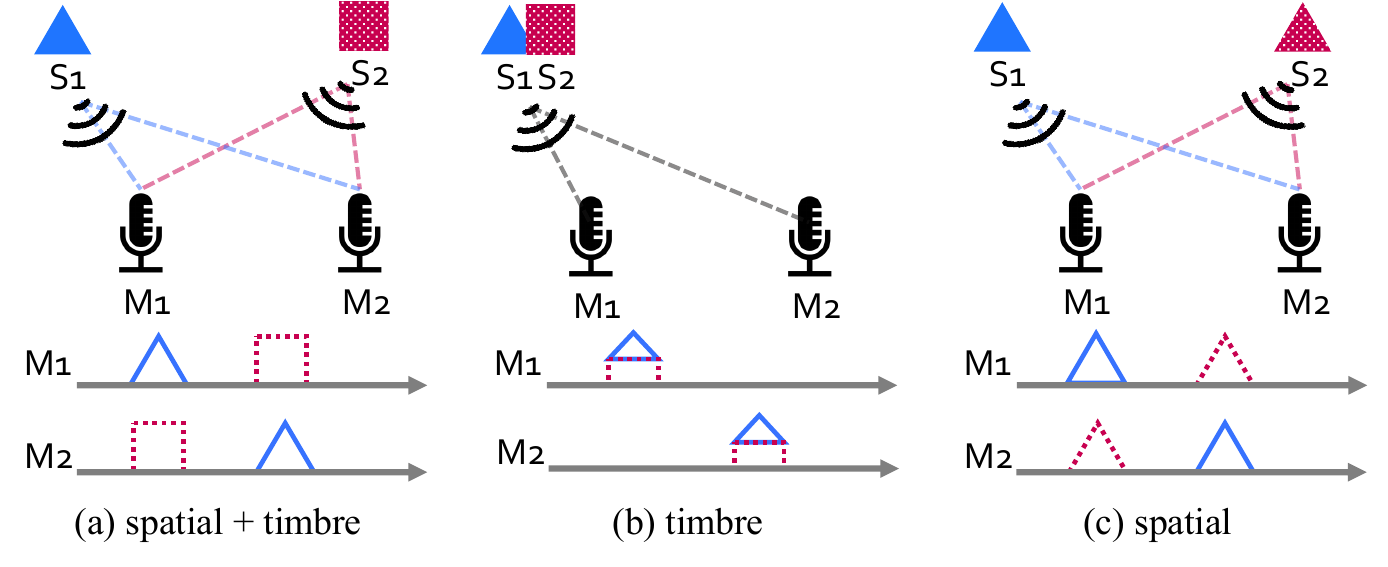}
    \caption{
        \emph{Illustrations of a multi-channel source separation dataset.}
        Subcaptions denote effective features to solve the task.
        Spatial feature is easy feature and timbre feature is hard feature.
        (a) Different sources (speakers) located at distinct positions, providing both timbre and spatial information for source separation; (b) Different sources co-located at the same position, offering only timbre information; and (c) The same source located at varying positions, yielding exclusively spatial information.
        We name this dataset SpatialVCTK, and we describe details in Section~\ref{subsec:spatialVCTK}.
    }
    \vspace{-0.5cm}
    \label{fig:multi_sep_setting}
\end{figure}

%% file: problem.tex
\section{Analysis of Feature Preference}
\label{sec:problem}

In this section, we formally describe the imbalanced feature preference,
and then provide an empirical analysis about it.

\subsection{Problem formulation}

The problem of imbalanced feature preference can occur when a targeted task is solvable in multiple ways utilizing different features.
But, it is more frequently realized, in particular, when the learning complexity
for each feature is severely different, c.f., simplicity bias \cite{simplicity} and shortcut learning \cite{minderer2020shortcut}.
For the sake of simplicity, we consider a supervised learning
given $\setDall$ with only two available features: easy and hard features.
Then, we can partition $\setDall{}= \setSeasy{} \cup \setShard{} \cup \setSboth{}$
into three disjoint subsets
such that
$\setSeasy{}$ (resp. $\setShard{}$) contains data solvable by only the easy feature (resp. the hard feature) and $\setSboth{}$ comprises data solvable by either of them. 
The core problem of imbalanced feature preference is that the model does not effectively learn the hard feature from $\setSboth$. This limits the generalization performance on tasks requiring the hard feature
even when $\setSboth$ carries sufficiently many samples on the hard feature.
Our goal is to obtain a balanced model capable of solving all partitions ($\setSeasy, \setShard, \setSboth$) 
without additional data collection.

\subsection{Multi-channel source separation}
\label{subsec:spatialVCTK}

Throughout this paper, 
    we consider 6-channel source separation, where
     two human speakers in a room 
     are recorded by a 6-channel microphone array 
    and we want to obtain signals from each speaker, separately. 
    To train a neural network for this, we obtain a dataset $\setDall$, called SpatialVCTK,
    based on the setup proposed in \cite{coneofsilence}
    using room acoustics simulator~\cite{pyroomacoustics}
    and VCTK dataset~\cite{VCTK} downsampled at sampling rate 16 kHz.
    In SpatialVCTK,
    the two sound sources are randomly located in the room
    where the angle between them at the microphone is uniformly at random. 
    The sources are uttered by 109 persons from VCTK~\cite{VCTK}, randomly selected. 
    
    In SpatialVCTK, we can correspond the spatial and timbre features,
    illustrated in Figure~\ref{fig:multi_sep_setting}, to the easy and hard features, respectively.
    To be specific, SpatialVCTK can be decomposed into 
      $\setSeasy$, $\setShard$, and  $\setSboth$ as follows.
      $\setSeasy$ is $0.9\%$ of $\setDall$, 
      consisting of samples where the two sources are from the same speakers, i.e., solvable by only the spatial features.
$\setShard$ is $11.1\%$ of $\setDall$
consisting of samples where the angle between two sources is less than 20$^{\circ}$, i.e., solvable by only the timbre feature.
The remaining $88.0\%$ of $\setDall$ is $\setSboth$, which is solvable by one of the two features.

\subsection{Preference to spatial feature}

\label{sec:exp-analysis}

Note that the full dataset $\setDall$ (SpatialVCTK) includes sufficient information to learn both timbre and spatial features. Indeed,
the largest subset $\setSboth$ contains various samples of the hard feature.
To analyze imbalanced feature preference in SpatialVCTK, 
we further obtain additional datasets $\setDeasy$ and $\setDhard$
which contains samples of only easy or hard feature, respectively.
In addition, for evaluation purpose, we use (unseen) test datasets $\setTall / \setTeasy / \setThard$ corresponding to training datasets $\setDall / \setDeasy / \setDhard$.

In Table~\ref{tab:problem}, we compare 
the source separation models trained on different datasets: $\setDall / \setDeasy / \setDhard$.
Training with $\setDall$ mostly learns the easy feature (spatial), and rarely learns the hard feature (timbre).
The model trained with $\setDall$ behaves similar to the model trained with $\setDeasy$. 
Although there is sufficient information about the hard feature (i.e., $\setSboth$, $\setShard$) in $\setDall$, the model exhibits a strong preference for learning easy features.
In other words, $\setSboth$ is used to learn only the easy feature.

\input{tables/table_problem}

\input{figures/figure_problem_forgetting}

    Moreover, Figure~\ref{fig:probelm_forgetting} exhibits this problem more severely.
    The model starts training in $\setDhard$ dataset and it learns the timbre feature. After changing the dataset into $\setDall$ dataset, the model ignores the timbre feature and learns the spatial feature.
    It means the $\setDall$, especially $\setSboth$, is used to learn easy feature, even if the model already has learned hard feature.

%% file: tables/table_problem.tex
\begin{table}[t]
\begin{minipage}{\linewidth}
    \caption{
        \emph{Comparison of separation models trained different datasets in SI-SDRi (dB).} 
        The size of each dataset is 5000. 
        }
    \label{tab:problem}
    \centering
        \vspace{-0.2cm}
    \begin{tabular}{c|ccc}
        \toprule
         & \multicolumn{3}{c}{training dataset} \\
        & $\setDall$ & $\setDhard$ & $\setDeasy$ \\ 
        \midrule
         $\setTall$  & 19.74	& 10.52	& 18.80       \\
        $\setThard$ & 0.16	& 11.32	&  -0.46   \\
        $\setTeasy$  & 19.32	& 5.72	& 18.89          \\
        \bottomrule
    \end{tabular}
    \vspace{-0.2cm}
\end{minipage}
\end{table}

%% file: figures/figure_problem_forgetting.tex
\begin{figure}[t]
    \centering
    \includegraphics[width=0.49\linewidth, trim={0cm, 0cm, 0cm, 0cm}]{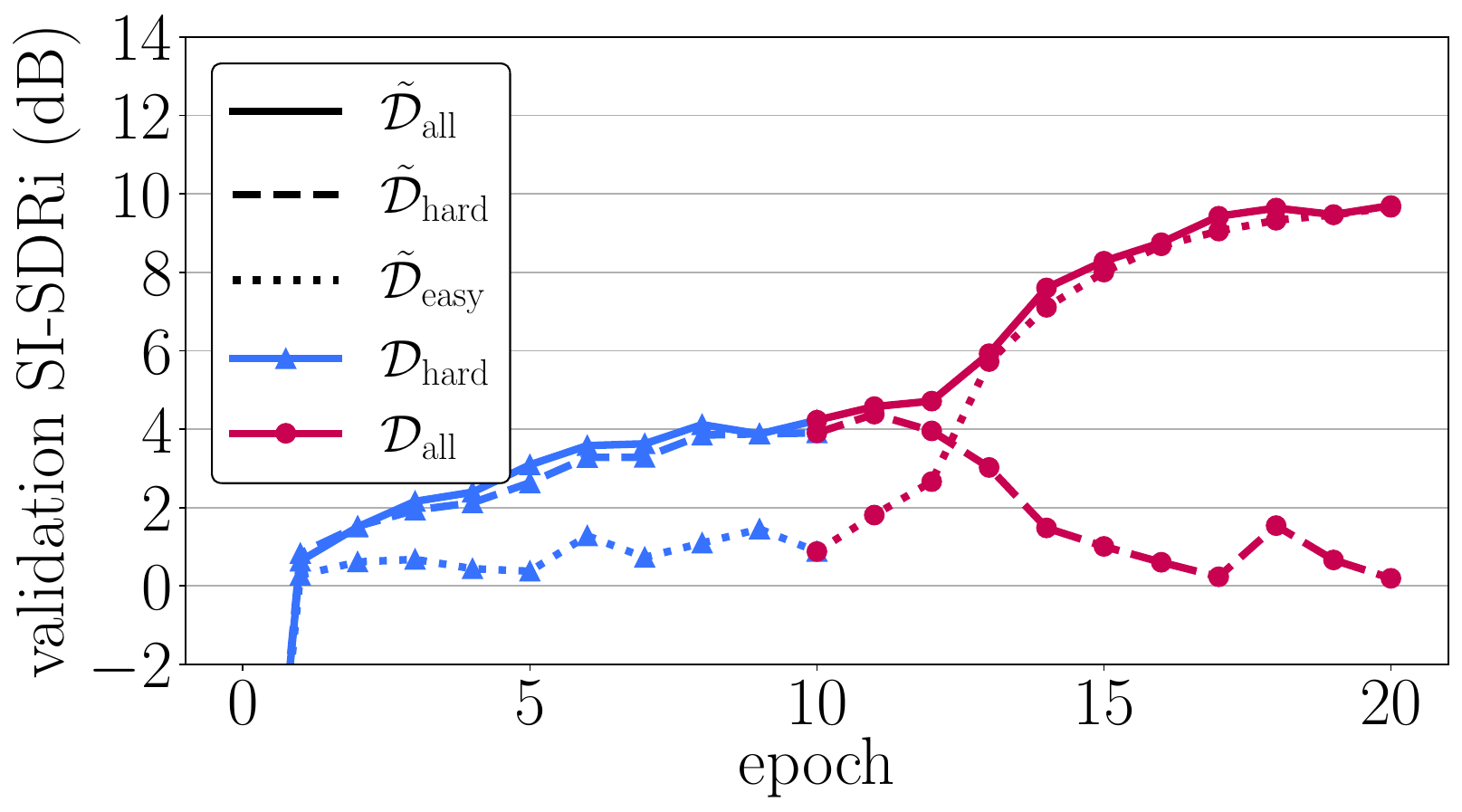}
    \includegraphics[width=0.49\linewidth, trim={0cm, 0cm, 0cm, 0cm}]{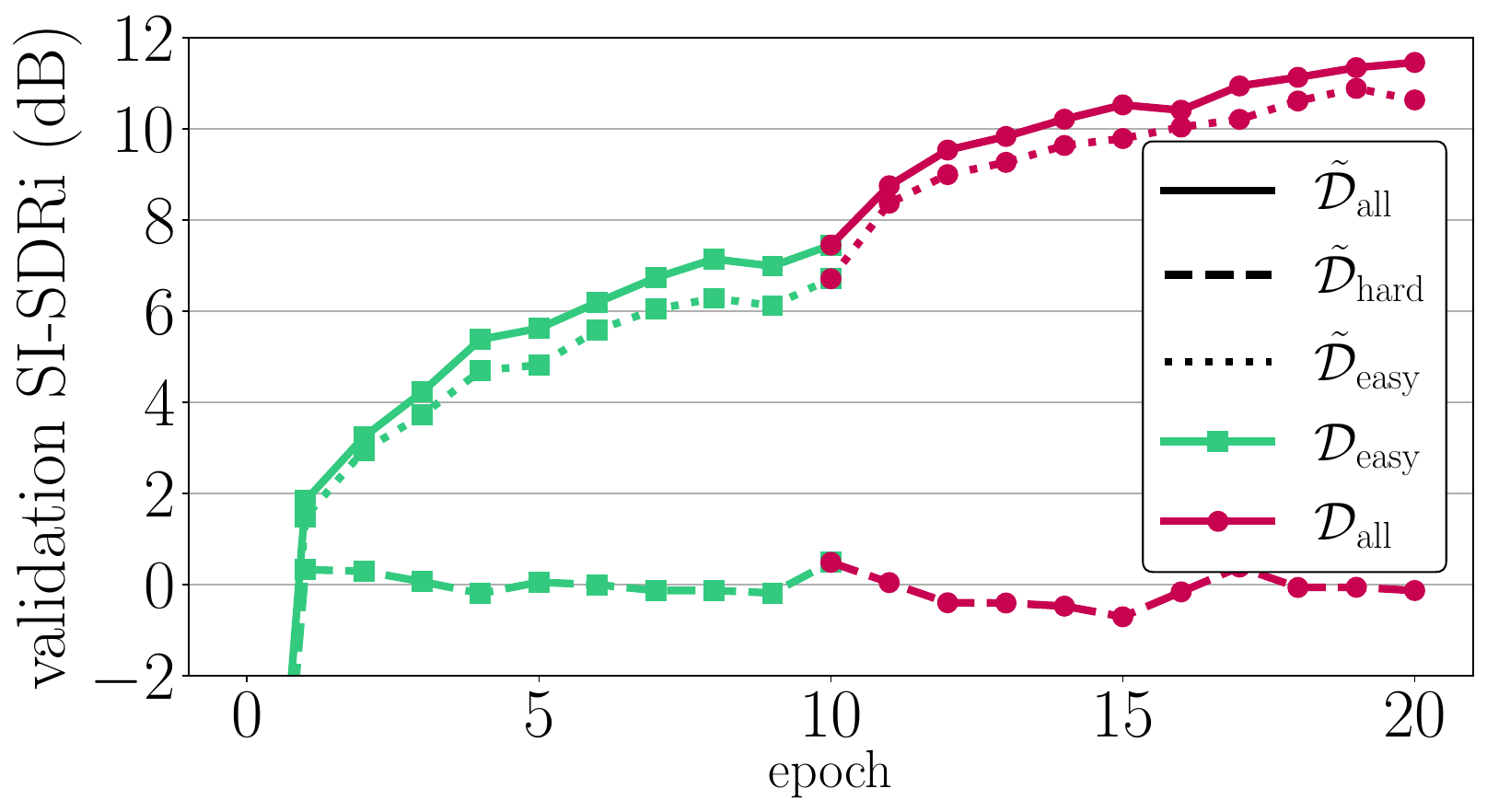}
    \vspace{-0.3cm}
    \caption{
       \emph{An example of forgetting the hard feature.}
       The color denotes training dataset.
       A model trained in $\setDhard$ forgets the hard feature after the training dataset is changed.
       Otherwise, a model trained in $\setDeasy$ does not forget the easy feature.
    }
    \vspace{-0.5cm}
    \label{fig:probelm_forgetting}
\end{figure}

%% file: method.tex
\section{Method}
\label{sec:method}

\input{figures/figure_method_overview}

We propose a method named FEAture BAlancing by Suppressing Easy feature (\ours). It facilitates the balanced learning of both easy and hard features from $\setDall$ containing the small $\setShard$.
\ours{} involves two steps:
\vspace{-1mm}
\begin{itemize}[leftmargin=0.3cm]
    \item Training a generative feature suppression model: A feature suppression model is designed to retain the hard feature from the given data while suppressing the easy feature. 
    We adopt WaveCycleGAN~\cite{tanaka2018wavecyclegan} to generate data only containing the hard feature (see Section~\ref{subsec:feature_suppression_model}). 
    \vspace{-1mm}
    \item Data augmentation via the feature suppression model: We utilize the trained feature suppression model to suppress easy features within $\setDall$ data, leaving only hard features. This process effectively increases the amount of data containing hard feature. To balance the augmented and original data, we use a reweighting method to ensure their appropriate contribution to the training process (see Section~\ref{subsec:augmentation}).
\end{itemize}
An overview of \ours{} is also illustrated in Figure~\ref{fig:method_overview}.

\vspace{-2mm}
\subsection{Generative feature suppression model}
\label{subsec:feature_suppression_model}

We obtain a generative feature suppression model based on the WaveCycleGAN~\cite{tanaka2018wavecyclegan} for multi-channel source separation.
To this end, we require four networks:
1) a forward generator that transforms source signals in $\setDall$ into those in the distribution of $\setShard$; 2) a forward discriminator that distinguishes between the transformed data and the actual ones in $\setShard$; 3) a reverse generator that reverts the transformed source signals into those in the distribution of $\setDall$, and 4) a reverse discriminator that differentiates between samples from the reverse generator and $\setDall$. 
Here, each generator takes two source signals as an input,
and transforms two source signals.
The forward generator is expected to be the feature suppression (FS) model,
which suppresses the easy feature of the source signals in $\setDall$. 
Then, this allows us to augment $\setShard$ from $\setDall$,
where the mixture signals are directly obtained by summing the source signals transformed by the feature suppression model.

\vspace{-1mm}
\subsection{Augmentation and reweighting for balanced learning}\label{subsec:augmentation}

The feature suppression model facilitates the training of the hard feature from the dataset $\setSboth$ by effectively suppressing the easy feature. 
However, achieving a balance in the training process between the easy and hard feature remains crucial. 
To enable the model to simultaneously learn both the easy and hard features, we employ a reweighting method.

In this approach, a percentage of $\lambda_\mathrm{aug}$ of the training data undergoes augmentation, while the remaining $1 - \lambda_\mathrm{aug}$ percent of the training data remains unaltered. This strategy ensures a proper balance between augmented and non-augmented data during the training phase, thereby promoting the effective learning of both easy and hard features.

%% file: figures/figure_method_overview.tex
\vspace{-3mm}
\begin{figure*}[!t]
    \centering
    \includegraphics[width=\linewidth]{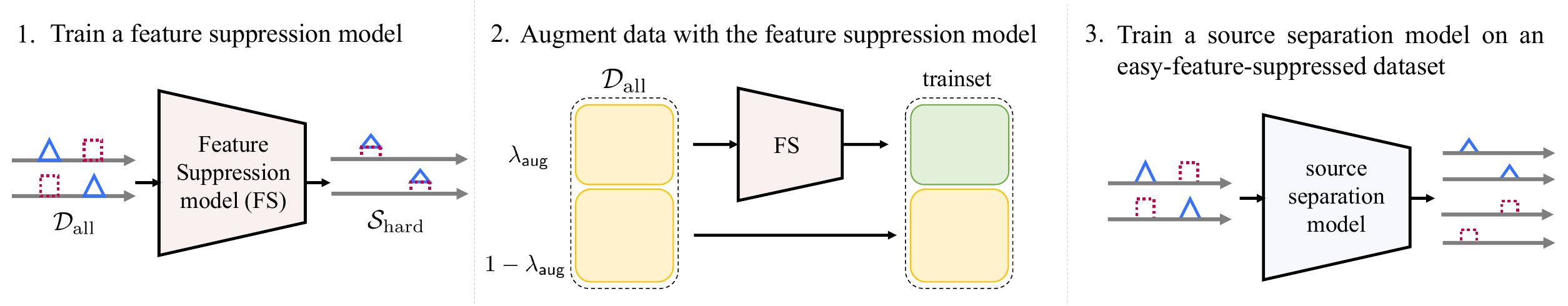}
    \vspace{-0.5cm}
    \caption{
       \emph{A schematic illustration of the proposed method (\texttt{FEABASE}) for training a source separation model.} 
    }
    \label{fig:method_overview}
    \vspace{-0.3cm}
\end{figure*}

%% file: experiments.tex
\vspace{-1mm}
\section{Experiment}
\vspace{-1mm}
\label{sec:experiment}

\subsection{Experiment settings}
\input{figures/figure_translation_qualitative}
\input{tables/table_exp_suppression}

    \paragraph{Evaluation metric}
        We employ an evaluation metric, Scale-Invariant Signal-to-Distortion-Ratio improvement (SI-SDRi), which is a variant of
        Scale-Invariant Signal-to-Distortion-Ratio (SI-SDR)~\cite{sdr}.
        SI-SDR compares original signal $s$ to predicted signal $\hat{s}$:
        \begin{align}
        \text{SI-SDR}(s, \hat{s}) =10 \log_{10} 
        \left( 
        {\big\| \tfrac{\hat{s}^T s}{\|s\|^2} s \big\|^2} 
        \middle/
        { \big\|\tfrac{\hat{s}^T s}{\|s\|^2} s - \hat{s} \big\|^2 }  \right) \;.
        \end{align}
        SI-SDRi is an improvement over SI-SDR with mixture $m$, i.e., $\text{SI-SDRi}(s, \hat{s}) = \text{SI-SDR}(s, \hat{s}) - \text{SI-SDR}(s, m)$.

    \paragraph{Source separation model}
        We modify a Dual-path RNN~\cite{dprnn} model for multi-channel audio.
        We call it \mdprnn.
        Compared to origianl Dual-path RNN, the input and output channels are expanded to be $6$.
        Loss function for the source separation task is SI-SDR with permutation invariant training~\cite{pit}.

     \paragraph{Feature suppression model}
        We use slightly changed vesion of WaveCycleGAN as the feature suppression model.
        We expand the input and output channel of WaveCyleGAN
        to 6 channels for multi-channel audio.

    \paragraph{Training details}
        For the \mdprnn{} model, we set the learning rate to 5e-4 and a batch size to 4. The training process continues for 100 epochs, and we reduce the learning rate by a factor of 0.5 when the validation error ceases to decrease.
        In the training process of feature suppression model, we employ a learning rate of $10^{-4}$, a batch size of 16, and conduct training for 50 epochs. The cycle consistency loss is weighted by a factor of 50, while the identity-mapping loss is weighted by a factor of 45.
        We train the feature suppression model with two datasets, 
        $\setDall$ and $\setShard$.
        The training dataset sizes for $\setDall$ and $\setShard$ are 5000 and 500, respectively.

    \paragraph{Baselines}
        Our baselines consist of \texttt{ERM}, \texttt{oracle re} and \texttt{oracle data}.
        \texttt{ERM} means training a model in its original form without any modification.
        \texttt{oracle re} trains a model 
        with a loss re-organized with
        different loss weights and sampling probabilities on $\setSeasy$, $\setShard$ and $\setSboth$, which are optimized by a grid search
        on $\{0.1, 0.3, 1, 3, 10, 30, 100, 300,$ $1000\}$.
        Note that \texttt{oracle re} is comparable to or better than reweighting and resampling methods~\cite{nam2020LfF, groupdro, jtt}. \texttt{oracle data} uses $\setDall$ and additional hard dataset of size $\{0, 500, 2000, 4500\}$.        
        This can be considered as a clear performance upper bound.

\vspace{-1mm}
\subsection{What feature suppression model learns}
\vspace{-1mm}

The feature suppression model learns to suppress easy (spatial) features of $\setDall$ but maintain hard features, resulting that each source sounds as if it is coming from the same direction.
Figure~\ref{fig:transfer_qualitative} illustrates the original data and feature suppressed data. In the original data, the arrival order of the sources is different for each channel. For example, source A arrives first at channel 0, while source B arrives first at channel 1. However, after feature suppression, both source A and source B arrive first at channel 0.
The corresponding audio samples are available at \url{http://ml.postech.ac.kr/feature-imbalance-separation/}.

Table~\ref{tab:exp_aug_result2} presents the quantitative evaluation of feature suppression and its impact on the separation models. The spatial separation model's performance significantly degraded after feature suppression, while the timbre separation model's performance only slightly degraded. It implies that the feature suppression model effectively suppresses spatial features while preserving timbre features. 
\vspace{-1mm}
\subsection{Comparison of \ours{} to baselines}
\vspace{-1mm}
\input{figures/figure_ablation_aug_ratio}

In this section, we evaluates the result of \ours{} method, in comparison with \texttt{ERM}, \texttt{oracle re}, and \texttt{oracle data}.
As shown in Figure~\ref{fig:ablation_aug_ratio}, there is a trade-off between learning timbre feature and learning spatial feature by different weight ratios. 
Maximum timbre SI-SDRi of \texttt{oracle re} is 7.14, because models are mainly trained using $\setShard$ of size 500. 
In contrast, \ours{} effectively enables the learning of hard feature, e.g., the maximum timbre SI-SDRi is 10.65, which is comparable to \texttt{oracle data}.
Also, not only learning hard feature, the Pareto front of \ours{} is far beyond the \texttt{oracle re}, except when the model mainly learns spatial feature.
This implies that our method successfully addresses the feature preference problem by achieving a more balanced performance for both features.

%% file: figures/figure_translation_qualitative.tex
\begin{figure}[t]
    \vspace{-0.2cm}

    \begin{minipage}{\linewidth}
    \centering
    \includegraphics[width=1\linewidth, trim={0cm, 0cm, 0cm, 0cm}]{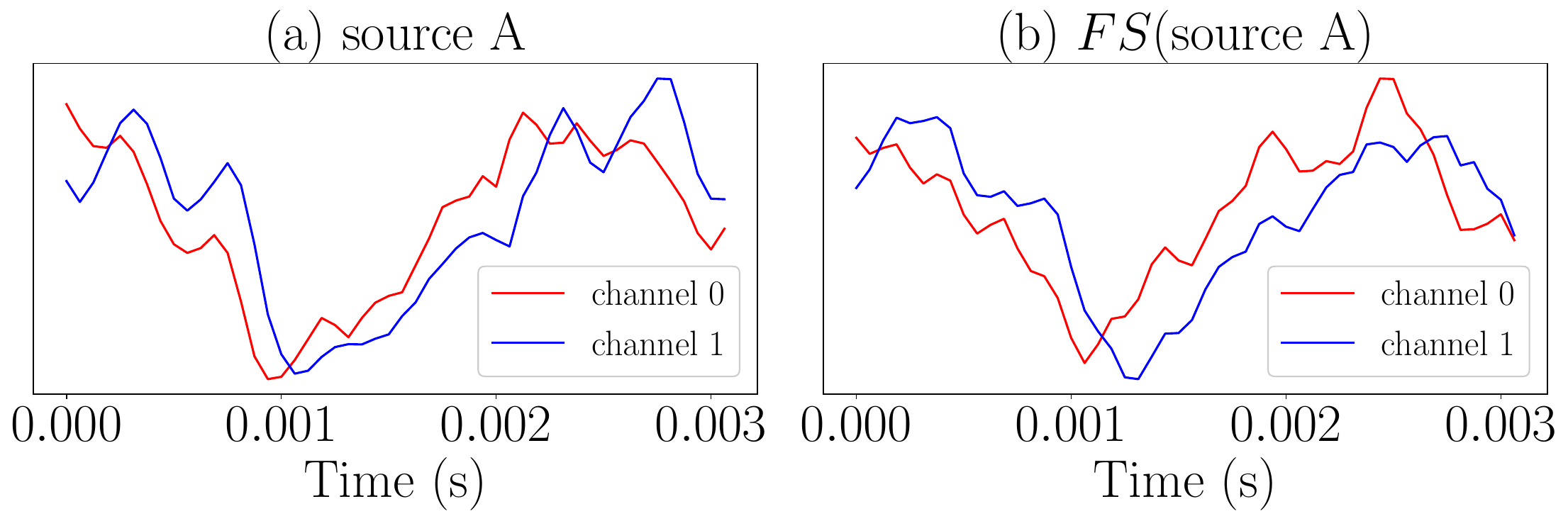}
    \includegraphics[width=1\linewidth, trim={0cm, 0cm, 0cm, 0cm}]{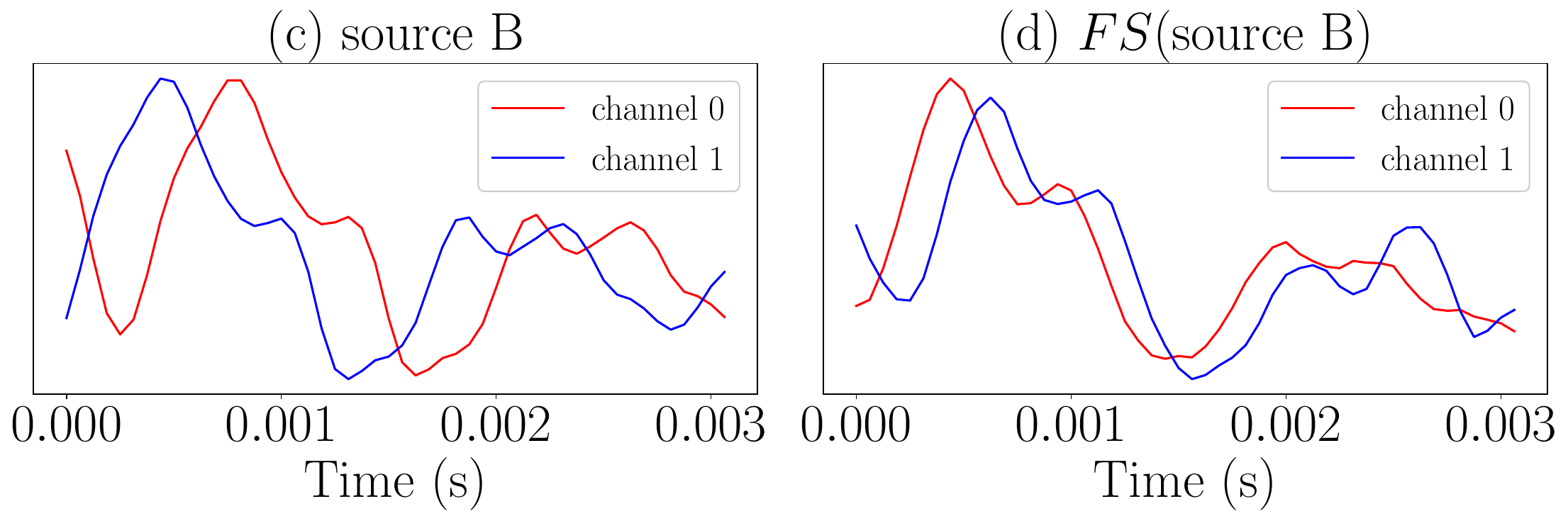}
    \vspace{-0.7cm}
    \caption{
       \emph{An example of feature suppression.}
       Red and Blue colors indicate audios of different channel.
       When channel 0 precedes channel 1, it indicates that the corresponding source is in proximity to channel 0's microphone.
       Original audios (left) have different direction, source A is close to channel 0 and source B is close to channel 1.
       Feature-suppressed audios (right) have similar spatial feature.
    }
    \label{fig:transfer_qualitative}
    \vspace{-0.5cm}
    \end{minipage}
\end{figure}

%% file: tables/table_exp_suppression.tex
\begin{table}[t]
    \caption{
        \emph{Changes of SI-SDRi (dB) by the feature suppression model.}
        $FS$ indicates the learned feature suppression model. 
        $FS$($\setTall$) refers to the $\setTall$ altered by the $FS$.
        $f_\mathrm{easy}$ refers to the model trained only with $\setDeasy$, while $f_\mathrm{hard}$ refers to the model trained only with $\setDhard$.
    }
    \label{tab:exp_aug_result2}
    \centering
    \vspace{-0.2cm}
    \resizebox{0.65\linewidth}{!}{
        \begin{tabular}{l|cc}
        \toprule
        & \multicolumn{2}{c}{model} \\
                 & $f_\textnormal{easy}$ & $f_\textnormal{hard}$ \\ \midrule
        before $FS$ : $\setTall$ & 19.32                    & 10.77                   \\ \midrule
        after $FS$ : $FS$($\setTall$)    & 2.45                     & 10.46           \\
        \bottomrule
        \end{tabular}
    }
    \vspace{-0.3cm}

\end{table}

%% file: figures/figure_ablation_aug_ratio.tex
\begin{figure}[t]
    \centering
    \includegraphics[width=0.9\linewidth, trim={0cm, 0cm, 0cm, 0.5cm}]{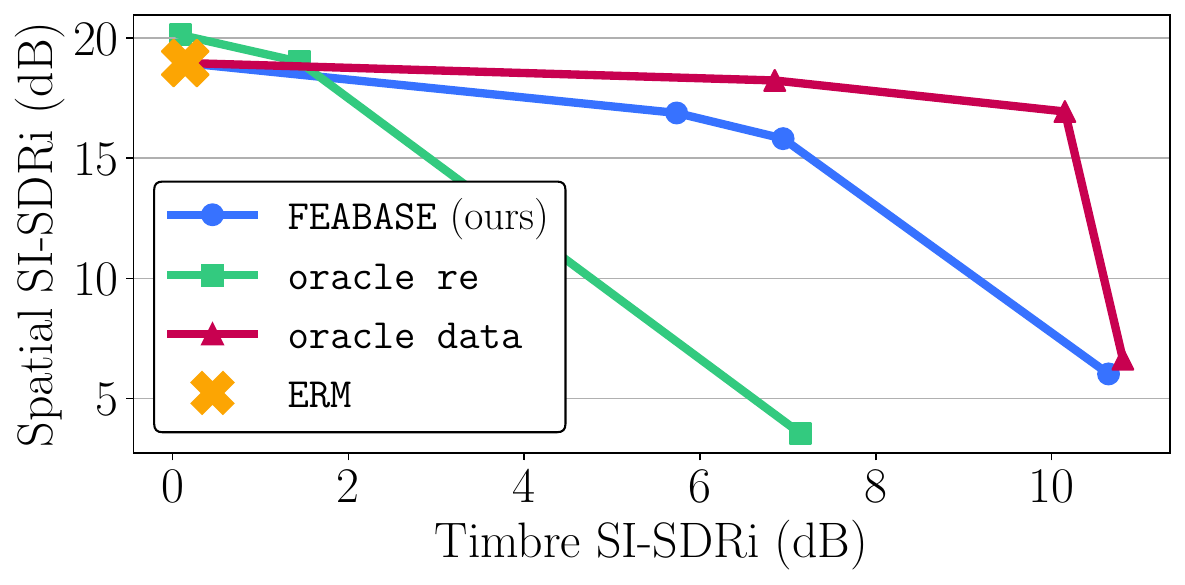}
    \vspace{-0.3cm}
    \caption{
       \emph{Pareto front of different methods.} 
       Different (augmentation, reweighting, data)  ratios are searched for \ours, {\tt oracle data}, and {\tt oracle re}. 
       \ours{} makes better Pareto front than {\tt oracle re}, by leveraging timbre feature significantly.
    }
    \vspace{-0.4cm}
    \label{fig:ablation_aug_ratio}
\end{figure}

%% file: main.bbl
\begin{thebibliography}{10}

\bibitem{simplicity}
Harshay Shah, Kaustav Tamuly, Aditi Raghunathan, Prateek Jain, and Praneeth Netrapalli,
\newblock ``The pitfalls of simplicity bias in neural networks,''
\newblock in {\em NeurIPS}, 2020.

\bibitem{minderer2020shortcut}
Matthias Minderer, Olivier Bachem, Neil Houlsby, and Michael Tschannen,
\newblock ``Automatic shortcut removal for self-supervised representation learning,''
\newblock in {\em ICML}, 2020.

\bibitem{nam2020LfF}
Junhyun Nam, Hyuntak Cha, Sungsoo Ahn, Jaeho Lee, and Jinwoo Shin,
\newblock ``Learning from failure: Training debiased classifier from biased classifier,''
\newblock in {\em NeurIPS}, 2020.

\bibitem{jtt}
Evan~Z Liu, Behzad Haghgoo, Annie~S Chen, Aditi Raghunathan, Pang~Wei Koh, Shiori Sagawa, Percy Liang, and Chelsea Finn,
\newblock ``Just train twice: Improving group robustness without training group information,''
\newblock in {\em ICML}, 2021.

\bibitem{kim2020m2m}
Jaehyung Kim, Jongheon Jeong, and Jinwoo Shin,
\newblock ``M2m: Imbalanced classification via major-to-minor translation,''
\newblock in {\em CVPR}, 2020, pp. 13896--13905.

\bibitem{coneofsilence}
Teerapat Jenrungrot, Vivek Jayaram, Steve Seitz, and Ira Kemelmacher-Shlizerman,
\newblock ``The cone of silence: Speech separation by localization,''
\newblock in {\em NeurIPS}. 2020, Curran Associates, Inc.

\bibitem{fasnettac}
Yi~Luo, Zhuo Chen, Nima Mesgarani, and Takuya Yoshioka,
\newblock ``End-to-end microphone permutation and number invariant multi-channel speech separation,''
\newblock in {\em ICASSP}, 2020.

\bibitem{doa}
Kohei Saijo and Robin Scheibler,
\newblock ``Spatial loss for unsupervised multi-channel source separation,''
\newblock in {\em INTERSPEECH}, 2022.

\bibitem{pyroomacoustics}
Robin Scheibler, Eric Bezzam, and Ivan Dokmanic,
\newblock ``Pyroomacoustics: A python package for audio room simulation and array processing algorithms,''
\newblock in {\em ICASSP}, 2018.

\bibitem{VCTK}
Christophe Veaux, Junichi Yamagishi, and Kirsten MacDonald,
\newblock ``Cstr vctk corpus: English multi-speaker corpus for cstr voice cloning toolkit,''
\newblock 2017.

\bibitem{tanaka2018wavecyclegan}
Kou Tanaka, Takuhiro Kaneko, Nobukatsu Hojo, and Hirokazu Kameoka,
\newblock ``Wavecyclegan: Synthetic-to-natural speech waveform conversion using cycle-consistent adversarial networks,''
\newblock in {\em IEEE Spoken Language Technology Workshop}, 2018.

\bibitem{sdr}
Jonathan~Le Roux, Scott Wisdom, Hakan Erdogan, and John~R. Hershey,
\newblock ``Sdr - half-baked or well done?,''
\newblock in {\em ICASSP}, 2019.

\bibitem{dprnn}
Yi~Luo, Zhuo Chen, and Takuya Yoshioka,
\newblock ``Dual-path rnn: efficient long sequence modeling for time-domain single-channel speech separation,''
\newblock in {\em ICASSP}, 2020.

\bibitem{pit}
Dong Yu, Morten Kolbæk, Zheng-Hua Tan, and Jesper Jensen,
\newblock ``Permutation invariant training of deep models for speaker-independent multi-talker speech separation,''
\newblock in {\em ICASSP}, 2017.

\bibitem{groupdro}
Shiori Sagawa, Pang~Wei Koh, Tatsunori~B Hashimoto, and Percy Liang,
\newblock ``Distributionally robust neural networks for group shifts: On the importance of regularization for worst-case generalization,''
\newblock {\em arXiv preprint arXiv:1911.08731}, 2019.

\end{thebibliography}
